\documentclass{ws-rv9x6}
\usepackage{subfigure}   
\usepackage{ws-rv-thm}   
\usepackage{ws-rv-van}   
\makeindex

\begin{document}

\chapter[Kinetics of Protein-DNA Interactions: First-Passage Analysis]{Kinetics of Protein-DNA Interactions: First-Passage Analysis}\label{ra_ch1}

\author[M. P. Kochugaeva et. al.]{Maria P. Kochugaeva, Alexey A. Shvets and Anatoly B. Kolomeisky\footnote{Corresponding author}}

\address{Department of Biomedical Engineering and System Biology Institute\\ Yale University\\ 
West Haven, CT, 06516, USA \\
maria.kochugaeva@yale.edu}

\address{Institute for Medical Engineering and Science, Massachusetts Institute of Technology, \\
Cambridge, MA 02142, USA \\
shvets@mit.edu}

\address{Department of Chemistry and Center for Theoretical Biological Physics,\\ Rice University, \\
Houston, Texas 77005, USA, \\
tolya@rice.edu}
\begin{abstract}
All living systems can function only far away from equilibrium, and for this reason chemical kinetic methods are critically important for uncovering the mechanisms of biological processes. Here we present a new theoretical method of investigating dynamics of protein-DNA interactions, which govern all major biological processes. It is based on a first-passage analysis of biochemical and biophysical transitions, and it provides a fully analytic description of the processes. Our approach is explained for the case  of a single protein searching for a specific binding site on DNA. In addition, the application of the method to investigations of the effect of DNA sequence heterogeneity, and the role multiple targets and traps  in the protein search dynamics are discussed.
\end{abstract}

\body


\section{Introduction}\label{ra_sec1}

One of the most striking features of living systems is their dynamic nature \cite{alberts,lodish}. Biological processes involve time-dependent fluxes of energy and materials, which makes them strongly deviating from the equilibrium. This implies that concepts of equilibrium thermodynamics have limited applications for biological systems, while the role of chemical kinetic methods that study the dynamical transformations is significant.\cite{phillips} In this chapter, we present a new theoretical method of investigating the complex mechanisms of chemical and biological phenomena, which is based on explicit calculations  of dynamic properties via a first-passage analysis. The first-passage ideas have been already widely utilized in studies of various processes in Chemistry, Physics and Biology.\cite{vankampen,redner} Our method employs these ideas in developing a discrete-state stochastic framework for analyzing the dynamics of protein-DNA interactions.

It is known that many biological processes start when  protein molecules bind to specific target sequences on DNA to initiate cascades of biochemical reactions.\cite{alberts,lodish}  This fundamental aspect of protein-DNA interactions has been studied extensively by various experimental and theoretical methods.\cite{halford04,mirny09,kolomeisky11} Although a significant progress in understanding the protein search phenomena has been achieved, the detailed molecular mechanisms remain not well clarified. Furthermore, there are extensive theoretical discussions on how to explain the fast dynamics of the protein search for the targets on DNA, which is also known as a facilitated diffusion.\cite{kolomeisky11}

Large amount of experimental observations, coming mostly from the single-molecule measurements, suggests that the protein search is a complex dynamic phenomenon which combines three-dimensional (in the bulk solution) and one-dimensional (on DNA) motions.\cite{halford04,mirny09,kolomeisky11}  But the most paradoxical observation is that the protein molecules spend most of the search time ($\ge$90-99$\%$) on the DNA chain where they diffuse very slowly, and they still find the targets very fast, in some cases faster than the bulk diffusion would allow.\cite{halford04,mirny09,kolomeisky11} Several theoretical ideas discussing the role of lowering of dimensionality, electrostatic effects, correlations between 3D and 1D motions, conformational transitions, bending fluctuations, and hydrodynamics effects have been proposed. However, the analysis shows that none of these mechanisms can fully explain the facilitated diffusion.\cite{veksler13} To investigate the mechanisms of protein-DNA interactions, we developed a discrete-state stochastic framework to take into account the most relevant chemical states and transitions in the system. The application of first-passage approach allows us to explicitly evaluate the dynamic properties and to clarify several important molecular aspects of protein-DNA interactions.

\section{Discrete-State Stochastic Model of Protein Search for the Specific Target on DNA}

\begin{figure}
\centerline{\includegraphics[width=8.8cm]{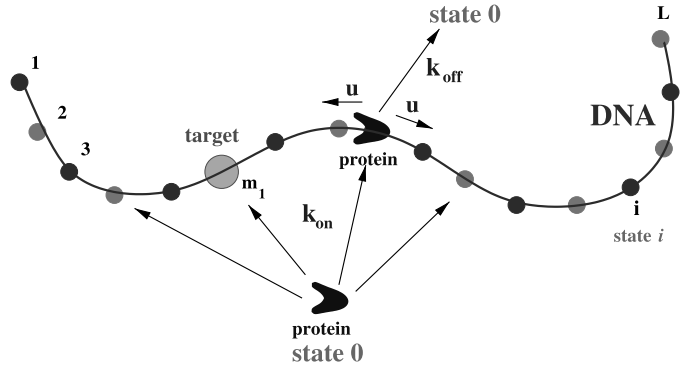}}
\caption{A schematic view of the discrete-state stochastic  model of the protein search. DNA chain has $L-1$ non-specific binding sites and one specific site. A protein molecule can diffuse along the DNA segment with a rate $u$ in both directions. It can also associate to  DNA  with a rate $k_{on}$ or dissociate with the a $k_{off}$.  The search is finished when the protein binds to the target site at the position $m$.}
\label{Fig1}
\end{figure}

To explain our method in detail, let us consider a simple model where one protein molecule diffuses through the bulk solution around the DNA molecule with occasional non-specific bindings to DNA (when the scanning along the chain is taking place) until the specific target site is located, as shown in Figure \ref{Fig1}.\cite{veksler13}
In the dsicrete-state model the DNA chain has $L$ binding sites, and one of them at the position $m$ is the target for the protein molecule. Because the diffusion of the proteins in the bulk is very fast, all solutions states for the protein are combined into one state that we label as a state $0$: see Figure \ref{Fig1}. It is assumed then that from the bulk solution the protein can bind with equal probability to any site on DNA, and the total association rate to DNA is equal to $k_{on}$, while the dissociation rate from DNA is $k_{off}$. The non-specifically bound proteins can diffuse along the DNA contour with a rate $u$ (see Figure 1).

Since the search process ends as soon as the protein molecule arrives to the specific site for the first time, we introduce a function $F_{n}(t)$, which is  defined a probability density function of reaching the site $m$ (target site) at time $t$ if at $t=0$ the protein started in the state $n$ ($n=0$ is the bulk solution, and $n=1,...,L$ are protein-DNA bound states). This function is also known as a first-passage probability density function.\cite{vankampen,redner} To determine these first-passage probabilities, we utilize backward master equations that describe the temporal evolution of these quantities,\cite{vankampen,redner,veksler13}
\begin{equation}\label{eq1}
\frac{F_{n}(t)}{d t}=u\left[ F_{n+1}(t)+F_{n-1}(t)\right] +k_{off} F_{0}(t) -(2u+k_{off}) F_{n}(t),
\end{equation}
for $2\le n \le L-1$, while at the boundaries ($n=1$ or $n=L$) we have
\begin{equation}\label{eq2}
\frac{F_{1}(t)}{d t}=u F_{2}(t) +k_{off} F_{0}(t) -(u+k_{off}) F_{1}(t),
\end{equation}
and
\begin{equation}\label{eq3}
\frac{F_{L}(t)}{d t}=u F_{L-1}(t) +k_{off} F_{0}(t) -(u+k_{off}) F_{L}(t).
\end{equation}
For the state $n=0$, the backward master equation is
\begin{equation}\label{eq4}
\frac{F_{0}(t)}{d t}=\frac{k_{on}}{L} \sum_{n=1}^{L} F_{n}(t) -k_{on} F_{n}(t).
\end{equation}
In the last equation, we used the fact that the rate to bind to any site on DNA is $k_{on}/L$, while the total association rate to DNA is $k_{on}$. In addition, the initial conditions require that $F_{m}(t)=\delta(t)$ and $F_{n \ne m}(t=0)=0$.

It is important to explain the physical meaning of the backward master equations because they differ from classical forward master equations widely employed in Chemical Kinetics. All trajectories that start at the state $n$ and end the target at the state $m$ can be divided into several groups. For example, for $2 \le n \le L$  all trajectories starting at $n$ will go to the state $n-1$, $n+1$ or to the state $0$ in the next time step, and the fraction of those trajectories is given by $u/(2u+k_{off})$,  $u/(2u+k_{off})$ and $k_{off}/(2u+k_{off})$, respectively. Equation (\ref{eq1}) describes this partition of the trajectories in the time-dependent manner. Thus the backward master equations reflect the temporal evolution of the first-passage probabilities.

The most convenient way to analyze the dynamics in the system is to use Laplace representations of the first-passage probability functions, $\widetilde{F_{n}(s)} \equiv \int_{0}^{\infty} e^{-st} F_{n}(t) dt$. Then Equations (\ref{eq1}),(\ref{eq2}), (\ref{eq3}) and (\ref{eq4}) can be written as simpler algebraic expressions:
\begin{equation}
(s+2u+k_{off}) \widetilde{F_{n}(s)}=u\left[ \widetilde{F_{n+1}(s)}+\widetilde{F_{n-1}(s)} \right]+k_{off} \widetilde{F_{0}(s)};
\end{equation}
\begin{equation}
(s+u+k_{off}) \widetilde{F_{1}(s)}=u\widetilde{F_{2}(s)}+k_{off} \widetilde{F_{0}(s)};
\end{equation}
\begin{equation}
(s+u+k_{off}) \widetilde{F_{L}(s)}=u\widetilde{F_{L-1}(s)}+k_{off} \widetilde{F_{0}(s)};
\end{equation}
\begin{equation}
(s+k_{on}) \widetilde{F_{0}(s)}=\frac{k_{on}}{L}\sum_{n=1}^{L}\widetilde{F_{n}(s)}.
\end{equation}
In addition, we have $\widetilde{F_{m}(s)}=1 $. These equations are solved assuming that the general form of the solution is $\widetilde{F_{n}(s)}=A y^{n} +B$, where the unknown coefficients $A$, $y$ and $B$ are determined from the initial and boundary conditions.\cite{veksler13} One could argue that the target site $m$ divides the DNA molecule into two homogeneous segments ($1\le n \le m$ and $m \le n \le L$), at which these parameters should have different values. It was shown\cite{veksler13} that this approach leads to explicit expressions for the first-passage probability functions. Specifically, one obtains
\begin{equation}
 \widetilde{F_{0}(s)}=\frac{k_{on}(k_{off}+s)S_{1}(s)}{L s(k_{off}+k_{on}+s)+k_{off}k_{on}S_{1}(s)}, 
\end{equation}
with an auxiliary function $S_{1}(s)$ defined as
\begin{equation}
    S_{1}(s)=\frac{y(1+y)(y^{-L}-y^{L})}{(1-y)(y^{1-m}+y^{m})(y^{m-L}+y^{1+L-m})};
\end{equation}
and with the parameters $y$ and $B$ given by
\begin{equation}\label{eq_y}
  y=\frac{s+2u+k_{off}-\sqrt{(s+2u+k_{off})^{2}-4u^{2}}}{2u};
\end{equation}
\begin{equation}
    B=\frac{k_{off}\widetilde{F_{0}(s)}}{(k_{off}+s)}.
\end{equation}

Explicit expressions for the first-passage probabilities provide a full dynamic description of the protein search processes. For example, the mean search time from the bulk solution, which is inversely proportional to the chemical association rate for the specific target site,  can be found from\cite{veksler13}
\begin{equation}\label{eq13}
    T_{0} \equiv -\frac{\partial F_{0}(s)}{\partial s} \bigg\vert_{s=0}=\frac{1}{k_{on}}\frac{L}{S_{1}(0)}+\frac{1}{k_{off}}\frac{L-S_{1}(0)}{S_{1}(0)}.
\end{equation}
This result has a clear physical meaning. Here the parameter $S_{1}(0)$ describes the average number of sites that the protein molecule scans during each visit to DNA while searching for a single specific site. Then, on average, to find the target the protein must make $L/S_{1}(0)$ visits to DNA. Each visit, on average, lasts $1/k_{on}$ while the protein scans for the target diffusing along the DNA chain. The protein also makes $L/S_{1}(0)-1$ dissociations back into the solution. The number of dissociation events is smaller by one than the number of association events because the last binding to DNA leads to finding the specific site.

\begin{figure}
\centerline{\includegraphics[width=8.8cm]{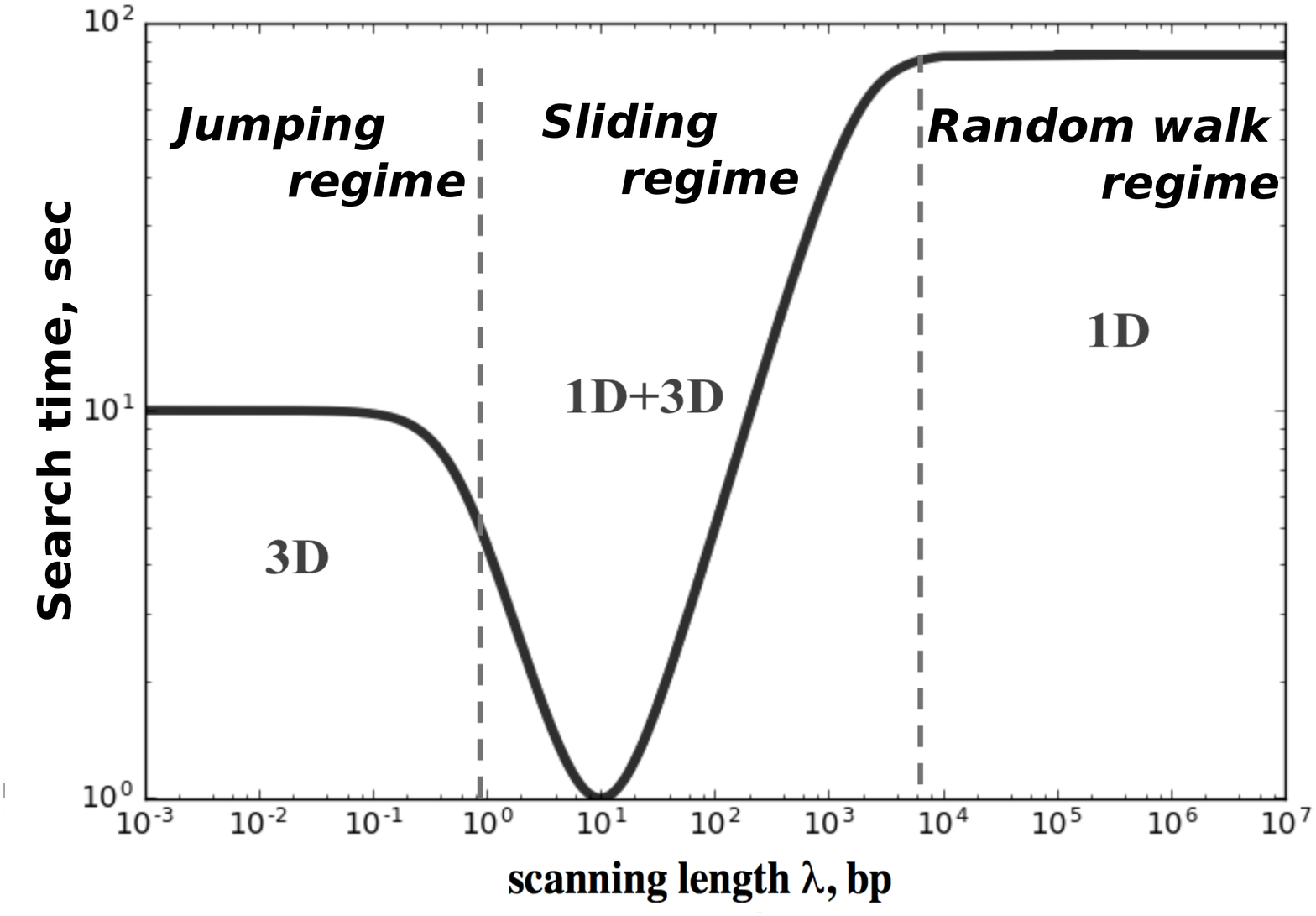}}
\caption{Average times to find the target on DNA for the protein molecule starting from the solution as a function of the scanning length $\lambda = \sqrt{u/k_{off}}$. The parameters are $L = 10^3$ bp, $u = k_{on} =10^5$ $s^{-1}$, and $m = L/2$. The transition rate $k_{off}$ is varied to change $\lambda$.}
\label{fig::phase_diag}
\end{figure}

The results of our calculations for the mean search times are presented in Figure \ref{fig::phase_diag}. Three dynamic search regimes are predicted depending on the values of kinetic parameters. If the protein molecule has a strong affinity to bind non-specifically the the DNA molecule (small $k_{off}$), then there will be only one searching cycle. After binding to DNA the protein will not dissociate until it finds the target. In this case, the mean search time scales $\sim L^{2}$ because the DNA-bound protein does a simple unbiased random walk. We call this dynamic phase a random walk regime. In the opposite limit of weak attractions between DNA and protein molecules (large $k_{off}$), the protein can bind to DNA but it cannot slide. So that the scanning length is of order of 1, and the protein makes $L$ searching cycles ($T_{0} \sim L$). This dynamic regime is called a jumping regime. The most interesting behavior is observed for intermediate interactions, which we label as a sliding regime. Here the scanning lengths are larger than one but smaller than the length of DNA $L$, and the number of searching cycles is also proportional to $L$. But in this regime the system can reach the most optimal dynamic behavior with the smallest search time. This search facilitation is achieved due to the fact that the fluxes to the target are coming from both the bulk solution and from the DNA chain.

\section{Multiple Targets and Traps}

The advantage of our method is the ability to extend and generalize it to more complex situations that can better describe the real biological systems. Let us present several examples to illustrate this. 

\begin{figure}
\centerline{\includegraphics[width=8.8cm]{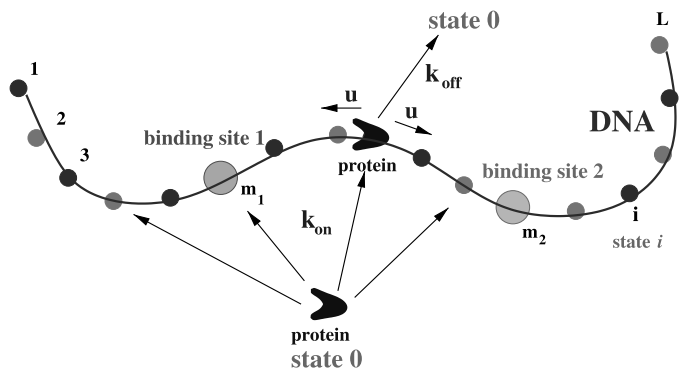}}
\caption{A discrete-state stochastic model of the protein search for multiple targets on DNA. The search ends when the protein finds one of the targets located at the sites $m_{1}$ and $m_{2}$. Adapted with permission from Ref. \cite{lange15a}}
\label{Fig3}
\end{figure}

In eukaryotic cells there are multiple target sites on accessible DNA fragments,\cite{lodish,alberts} which raises a question how long does it take for a protein to find {\it any} specific binding site on DNA. It has been argued that the mean search time in this system might not decrease proportionally to the number of targets, as one would naively expect, due to the complex mechanism of the search that involves both 3D and 1D motions.\cite{lange15a} So we extended our discrete-state stochastic framework to consider a model with multiple targets as presented in Figure \ref{Fig3}. To describe the search dynamics in this system, we again introduce a first-passage probability function $F_{n}(t)$ of finding {\it any} of the target at time $t$ if the process started at $t=0$ at the site $n$. Solving the corresponding backward master equations leads to the explicit expression for the mean search time for any number of targets,\cite{lange15a}
\begin{equation}
    T_{0}=\frac{1}{k_{on}}\frac{L}{S_{i}(0)}+\frac{1}{k_{off}}\frac{L-S_{i}(0)}{S{i}(0)},
\end{equation}
with the function $S_{i}(0)$ describing the average number of scanned sites on DNA with $i$ targets. This formula clearly is a generalization of Equation (\ref{eq13}) when there is only one target  ($i=1$). Specific expressions for $S_{i}(0)$ for various numbers of randomly distributed targets have been obtained.\cite{lange15a} For example, for $i=2$ it was shown that
\begin{equation}\label{eqS2}
    S_{2}=\frac{(1+y)\left[2(1-y^{2L+m_{1}-m_{2}})+(1-y^{m_{2}-m_{1}})(y^{2m_{1}-1}+y^{1+2(L-m_{2})})\right]}{(1-y)(1+y^{2m_{1}-1})(1+y^{1+2(L-m_{2}}))(1+y^{m_{2}-m_{1}})},
\end{equation}
where the parameter $y$ is given in Equation \ref{eq_y}.

\begin{figure}[htbp]
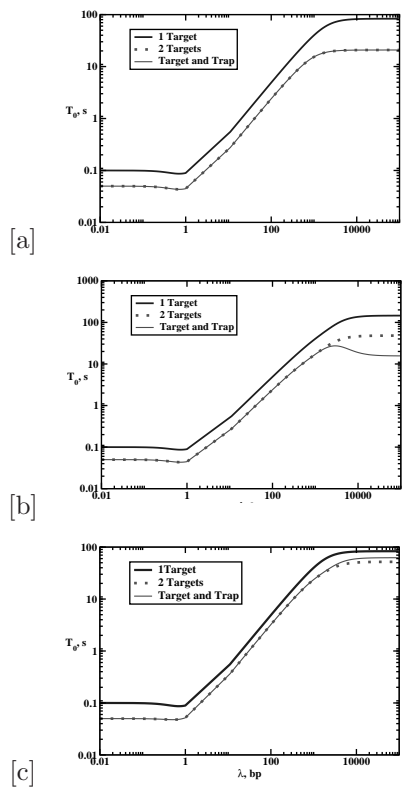

	\centering
	[a] \hskip 0.1 in \includegraphics[scale=0.30]{figures/FigTTa.eps} \vskip 0.1in
    [b] \hskip 0.1 in \includegraphics[scale=0.30]{figures/FigTTb.eps} \vskip 0.1in
    [c] \hskip 0.1 in \includegraphics[scale=0.30]{figures/FigTTc.eps} \vskip 0.1in
	\caption{Dynamic phase diagrams for the protein search on DNA with one target at the position $m$, with two targets at the positions $m_{1}$ and $m_{2}$ and with the target and the trap  at the positions $m_{1}$ and $m_{2}$. Parameters used for calculations are: $k_{on}=u=10^5$  s$^{-1}$ and $L=10000$. a) $m=L/2$, $m_{1}=L/4$ and $m_{2}=3L/4$; b) $m=L/4$, $m_{1}=L/4$ and $m_{2}=L/2$; and c) $m=L/2$, $m_{1}=L/2$ and $m_{2}=L$. Adapted with permission from Ref. \cite{lange15b}}\label{fig4}
\end{figure}

The results of explicit calculations for the mean search times are presented in Figure {\ref{fig4}. The presence of multiple targets does not affect the overall dynamic phase diagram: three search regimes are observed depending on the size of the scanning length and the size of the DNA segment, and in most cases the search is faster.  However, increasing the number of specific sites  might not always accelerate the search. To quantify this effect, we introduced an acceleration parameter, $a_{n}=T_{0}(1)/T_{0}(n)$, where $T_{0}(n)$ is the mean search for the system with $n$ targets. This ratio gives a numerical value of how faster the search in the presence of $n$ targets in comparison with a single-target system. It is interesting to analyze the results given in Figure \ref{Fig5}. One can see that there is a range of parameters when, surprisingly, the search dynamics in the system with two targets can be slower than the dynamics in the system with one target. This happens in the effectively 1D search regime  when the single target is located in the middle of the DNA chain, while two targets are close to each other and located near one of the ends of the DNA segment. Our theoretical analysis predicts that the degree of acceleration due to the presence of multiple targets depends on the nature of the dynamic search phase and on the location of the specific sites.\cite{lange15a}

\begin{figure}
\centerline{\includegraphics[width=8.5cm]{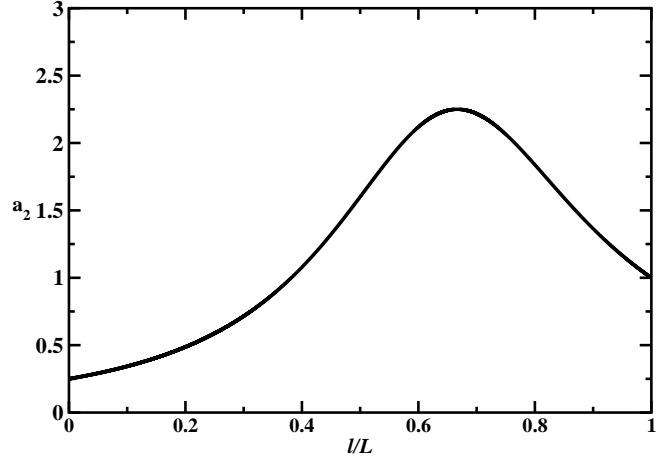}}
\caption{Ratio of the mean search times as a function of the normalized distance between the targets for single-target and two-target systems ($l$ is the distance between between targets, $L$ is the DNA length). The single target is in the middle of the chain. In the two-target system, one of the specific sites is fixed at the end and the position of the second one is varied. The parameters used in calculations are: $u=k_{on}=10^{6}$ s$^{-1}$; $k_{off}=10^{-4}$ s$^{-1}$; and $L=10000$. Adapted with permission from Ref. \cite{lange15a}}
\label{Fig5}
\end{figure}

Another important factor in the protein search is the existence of DNA sequences that have chemical compositions similar to the specific targets. Because the protein molecule can be trapped in these sites, the dynamics might be strongly affected. To analyze this effect, we extended the original model to include the possibility of traps, assuming that associations to these semi-specific sites are irreversible.\cite{lange15b} This assumption is reasonable because the search times in many systems are  relatively short and the experimental observations also have limited duration times. Our first-passage method can be easily applied here, but we have to notice that only a fraction of trajectories will reach the target site. Then the main quantity of our calculations, the first-passage probability function $F_{n}(t)$, is now a {\it conditional} probability for the protein molecules  not captured by the trap to find the target site. 

Let us consider a system consisting of a single target at the site $m_{1}$ and a single trap at the site $m_{2}$ on the DNA molecule with $L$ sites.\cite{lange15b} Solving the corresponding backward master equations yields the Laplace transform of the first-passage probability function to find the target if the protein starts from the bulk solution,\cite{lange15b}
\begin{equation}
\widetilde{F_{0}(s)}=\frac{k_{on}(k_{off}+s)S_{0}(s)}{L s(k_{off}+k_{on}+s)+k_{off}k_{on}S_{2}(s)}, \end{equation}
with 
\begin{equation}
    S_{0}(s)=\frac{(1+y)(1-y^{m_{1}+m_{2}-1})}{(1-y)(1+y^{2m_{1}-1})(1+y^{m_{1}-m_{2}})},
\end{equation}
and the parameters $y$ and $S_{2}$ given in Equations (\ref{eq_y}) and (\ref{eqS2}), respectively. This allows us to evaluate all dynamic properties in the system. The probability to reach the target (the fraction of successful trajectories) is given by the so-called splitting probability function,
\begin{equation}
    \Pi=\widetilde{F_{0}(s=0)}=\frac{S_{0}(0)}{S_{2}(0)}.
\end{equation}
The mean search time, which is the conditional mean first-passage time to reach the target, can be estimated by averaging over the successful trajectories, producing
\begin{equation}
T_0 \equiv -\frac{\frac{\partial \widetilde{F_0}(s)}{\partial s}\bigg\vert_{s=0}}{\Pi}=\frac{1}{k_{on}}\frac{L}{S_{2}(0)}+\frac{1}{k_{off}}\frac{L-S_{2}(0)}{S_{2}(0)} +\Pi \frac{\partial}{\partial s} \left[ \frac{S_{2}(s)}{S_{0}(s)} \right] \bigg\vert_{s=0}.
\end{equation}
It is interesting to note that the first two terms in this expression is exactly the mean search time for the system with two targets (at the sites $m_{1}$ and $m_{2}$) as we discussed above,\cite{lange15a} while the third term is a correction which accounts for the fact that the site at $m_{2}$ is a trap. The main reason for this is the observation that the sites $m_{1}$ and $m_{2}$ are special locations where all trajectories are end up in both systems, with two targets and with the target and the trap. For the two-target case the mean search times are averaged over all trajectories, while for the target and the trap system the mean search times are obtained only by considering the trajectories finishing at the target.\cite{lange15b}

Theoretical calculations for dynamic properties of the protein search in the presence of traps are illustrated in Figures \ref{fig4} and \ref{fig6}. Again three dynamic search phases are predicted, but adding the trap generally facilitates the search dynamics: see Figure \ref{fig4}. However, this acceleration (in comparison with the single-target system) is associated with lowering of the probability of reaching the specific target, as shown in Figure \ref{fig6}. In addition, the search dynamics depends on the nature of the dynamic phase. The strongest effect is observed in the random-walk regime (because it has only one searching cycle) where the locations of the target and the trap strongly influence the search.

\begin{figure} 
	\centerline{
	\includegraphics[width=8.5cm]{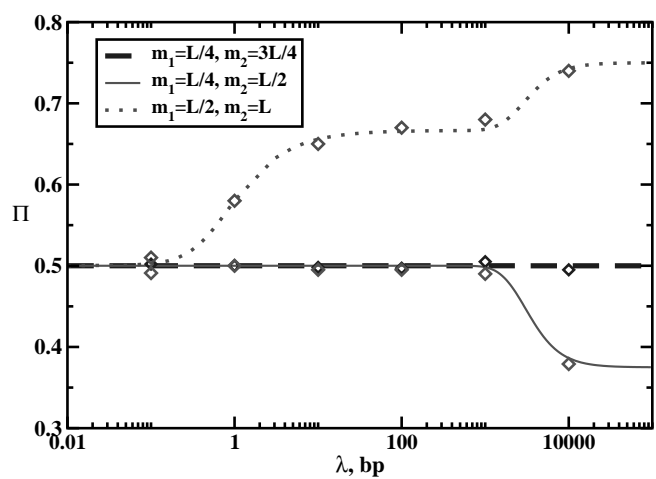}} 
        \caption{Probability to reach the target as a function of the scanning length for different distributions of the target and trap sites. Parameters used for calculations are: $k_{on}=u=10^5$  s$^{-1}$, $L=10000$ and $k_{off}$ is changing. Symbols are from Monte Carlo computer simulations. Adapted with permission from Ref. \cite{lange15b}}\label{fig6}
\end{figure}

\section{Sequence Heterogeneity}

It is known that real DNA  are heterogeneous polymer molecules consisting of several types of subunits, and the interactions between protein and DNA molecules depend on the DNA sequence at the interaction site. Obviously, this should affect the protein search dynamics because the diffusion rate for the non-specifically bound proteins must be position-dependent. Our theoretical framework is a convenient tool to investigate this problem.\cite{shvets15}

To model the role of DNA sequence heterogeneity on protein search dynamics, we assume a simplified picture in which each monomer in DNA  can be in one of two chemical species, $A$ or $B$.\cite{shvets15} When the protein is bound to the subunit $A$ ($B$), it interacts with energy $\varepsilon_A$ ($\varepsilon_B$), and the difference between interaction energies is given by $\varepsilon = \varepsilon_A - \varepsilon_B \ge 0$. This means that the protein attracts stronger to the $B$ sites than to the $A$ sites. The protein molecule can diffuse along DNA with a rate $u_A \equiv u$ or $u_B=ue^{-\varepsilon}$, where $\varepsilon$ is measured in $k_{B}T$ units. In addition, we assume that, independently of the chemical nature of neighboring sites, sliding out of the sites $A$ is characterized by the rate $u_A$, while the diffusion out of the sites $B$ is given by $u_B$. From the solution the protein associates with any site $A$ or $B$ on DNA with the corresponding rates $k^A_{on}=k_{on}$ or $k^B_{on}=k_{on}e^{-\theta\varepsilon}$. Note that for convenience the on-rates defined here as the rates per unit site, in contrast to our definitions in the previous sections. Similarly, the dissociations from the DNA chain are described by the rates $k^A_{off}=k_{off}$ and $k^B_{off}=k_{off}e^{(\theta-1)\varepsilon}$. Here, the parameter $0\le \theta\le1 $ specifies how the protein-DNA interaction energy is distributed between the association and dissociation transitions.\cite{shvets15} The physical meaning of this parameter is that the protein molecule tends to bind faster and to dissociate slower from the stronger attracting sites $B$, as compared with weaker attracting $A$ sites. The parameter $\theta$ accounts for these effects. To quantify the role of sequence heterogeneity, we consider the DNA molecule with a fixed chemical composition (the fractions of $A$ and $B$ monomers are the same), but with different arrangements of subunits. Two limiting cases are specifically analyzed. One of them views the DNA molecule as two homogeneous segments of only $A$ and only $B$ subunits separated by the target in the middle of the chain. Another one is the DNA chain with the alternating $A$ and $B$ sites. The block copolymer has a more homogeneous sequence, while the alternating polymers are more heterogeneous. It is important to note that in both cases, the overall interaction between the protein and DNA is the same (because the overall chemical composition in both cases is identical), and thus our analysis probes only the effect of the heterogeneity.

\begin{figure}
\centerline{\includegraphics[width=8.5cm]{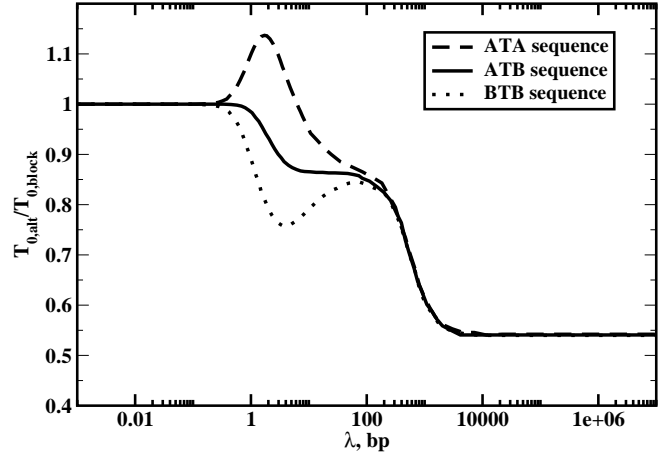}}
\caption{The ratio of the mean search times for the alternating DNA sequences and for the block copolymer DNA sequences as a function of the scanning length $\lambda=\sqrt{u/k_{off}}$. Three different chemical compositions near the target ($T$) are distinguished, namely, $ATA$, $ATB$, $BTB$. The transition rates are $u = 10^5\, s^{-1}$ and $k_{on} = 0.1\, s^{-1}$. The DNA length is $L = 1000$, the loading parameter is $\theta = 0.5$, and the energy difference of interactions for the protein with $A$ and $B$ sites is $\varepsilon=5$ $k_{B}T$. Adapted with permission from Ref. \cite{shvets15}}
\label{fig::seq_rel_time}
\end{figure}

Applying the first-passage approach to calculate the dynamic properties in this system leads to the explicit expressions for mean search times for all situations.\cite{shvets15} For example, for the block copolymer sequences, we obtain
\begin{equation}
T_{0}=\frac{k_{off}+k_{on}\left[(L/2-P_{A})+e^{\varepsilon}(L/2-P_{B})\right]}{k_{on} k_{off}(1+P_{A}+e^{\theta \varepsilon} P_{B})},
\end{equation}
where
\begin{equation}
    P_{i}=\frac{x_{i}^{1-L/2}-x_{i}^{1+L/2}}{(1-x_{i})(x_{i}^{1+L/2}+x_{i}^{_L/2})},
\end{equation}
\begin{equation}
x_{i}=\frac{2u_{i}+k_{off}^{(i)}-\sqrt{(2u_{i}+k_{off}^{(i)})^{2}-4u_{i}^{2}}}{2u_{i}},
\end{equation}
for $i=A$ or $B$. The expressions for the mean search time for alternating sequences are more bulkier and they can be found in Ref. \cite{shvets15}  The results of our calculations are presented in Figure \ref{fig::seq_rel_time} where the ratio of the mean search times for the block copolymer and alternating sequences are plotted.

One can see that the effect of the sequence heterogeneity on protein search dynamics depends on the nature of the dynamic phase. In the jumping regime when the protein does not slide along the DNA contour ($\lambda <1$, where the parameter $\lambda$ is proportional to the number of visited sites during each binding to DNA), the symmetry of the sequence does not play any role. This is because in this case, the process is taking place only via associations and dissociations (3D search), and the structure of the DNA chain is not important. The situation is different for the intermediate sliding regime (3D+1D search, $1 < \lambda < L$) where in most cases, the search on alternating sequences is faster. This can be explained by noticing that the search time in this dynamic phase is proportional to $L/\lambda$, which gives the average number of cycles before the protein can find the target. In the block copolymer sequence, the protein mostly comes to the target from the $B$ segment because of stronger interactions with these sites. In the alternating sequences, the protein can reach the target from both sides of DNA, and this lowers the overall search time. It can be shown analytically that the scanning length on the alternating segment is larger than the scanning length for the $B$ segment, i.e., $\lambda_{AB} > \lambda_B$.\cite{shvets15} Then the search is faster for the alternating sequence because $L/\lambda_{AB} < L/\lambda_B$, i.e., the number of searching cycles is lower for the alternating sequences. The only deviation from this picture is found in $ATA$ sequences where for the small range of parameters the search is slower than in the block copolymer sequence. This effect can be explained by the fact that the protein does not sit at $A$ sites for the long time and it moves quickly away, effectively increasing the barrier to enter the target via DNA. \cite{shvets15}

In the random-walk regime (1D search, $\lambda > L$), the effect of the sequence heterogeneity is even stronger: the protein molecule finds the specific binding site up to 2 times faster for more heterogeneous alternating DNA sequences. To understand this behavior, we note that in this case the mean first-passage time to reach the target is a sum of residence times on the DNA sites since the protein will not dissociate until the target is located. Because the target is in the middle of the chain, the mean time to reach the target from the block copolymer sequence will be $T_0 \simeq (L/4) \tau_B$, where $\tau_B$ is the average residence time at any site B. The protein prefers to start the search at any position on the $B$ segment with equal probability, i.e., the distance to the target varies from 0 to $L/2$. Then, the average starting position of the protein is $L/4$ sites away from the target. For the alternating sequences, the average distance to the target is approximately the same ($L/4$), but the chemical composition of intermediate sites on the path to the target is different, yielding, $T_0 \simeq (L/8) \tau_A + (L/8) \tau_B$ ($\tau_{A}$ is the residence time on $A$ sites). The protein spends much less time on $A$ subunits, and this leads to faster search for the alternating DNA sequences. For $\tau_A \ll \tau_B$, this also explains the factor of 2 in the search speed. In this case, the $B$ subunits can be viewed as effective traps that slow down the search dynamics. Thus, our theoretical calculations make surprising predictions that the sequence heterogeneity almost always accelerates the protein search for targets on DNA. And the stronger the contribution of 1D search modes, the stronger will be the effect of the sequence heterogeneity.

\section{Other Problems}

In addition to problems discussed above, we extended and generalized the first-passage method to a variety of problems associated with protein-DNA interactions. More specifically, the role of crowding on DNA during the protein search was explicitly investigated.\cite{shvets16} It was found that the mobility of crowding agents (other DNA bound proteins) is a key factor affecting the facilitated diffusion: highly mobile crowders do not affect the search, while slow crowders inhibit the search dynamics.  Similar analysis have been done for the protein search in the presence of static and dynamic obstacles (particles that occupy specific sites on DNA).\cite{shvets15b}  In this case, it was found that the key properties determining the search dynamics are size of the obstacle, the distance between the obstacles and the target, the dynamics of the obstacle and the nature of the dynamic search phase. We also analyzed the role of DNA looping for the target search by multisite proteins.\cite{shvets16b} In addition, the effect of protein conformational transitions on the target search has been also explored.\cite{kochugaeva16} Furthermore, we studied the surface-assisted dynamic processes by generalizing our method to 2D surfaces.\cite{shin18} It is also important to note that our theoretical method has been successfully applied for analyzing of several important experimental observations, including the kinetic studies of inducible transcription factor Egr-1,\cite{esadze14} the mechanism of the homology search by RecA protein filaments,\cite{kochugaeva17} and the dynamics of genome interrogation by CRISPS-Cas9 protein-RNA systems.\cite{shvets17}

\section{Conclusions}

A new theoretical method to investigate the dynamics of protein-DNA interactions is presented. It utilizes a discrete-state stochastic framework where dynamic properties are explicitly evaluated using the first-passage approach. Our approach takes into account the most relevant biochemical states and transitions. Because of exact calculations, it allows us to clarify many features of the complex mechanisms in these biological systems. The method is also successfully applied for understanding various experimental results. Thus, our theoretical approach is a powerful tool in studying protein-DNA interactions. There are several directions that we are planning to follow in the future studies.  They include the role of coupling between 1D, 2D and 3D motions, the effect of crowing in the bulk solutions, heterogeneity of the transition rates, DNA topological effects and many others.

\section*{Acknowledgments}
A.B.K. acknowledges the support from the Welch Foundation (C-1559), from the NSF (CHE-1664218) and from the Center for Theoretical Biological Physics sponsored by the NSF (PHY-1427654).

\bibliographystyle{ws-rv-van}

\begin{thebibliography}{99}

\bibitem{alberts} B. Alberts et al., {\it Molecular Biology of Cell}, 6th ed., Garland Science, New York (2014).

\bibitem{lodish} H. Lodish et al., {\it Molecular Cell Biology}, 6th ed., W. H. Freeman, New York (2007).

\bibitem{phillips} R. Phillips, J. Kondev, and J. Theriot, {\it Physical Biology of the Cell}, 2nd ed., Garland Science, New York (2012).

\bibitem{vankampen} N.G. Van Kampen, {\it Stochastic Processes in Physics and Chemistry}, 3rd ed., North Holland, Amsterdam (2007).

\bibitem{redner} S. Redner,{\it A Guide to First-Passage Processes}, Cambridge University Press, Cambridge (2001).


\bibitem{halford04} S.E. Halford and J.F. Marko, How do site-specific DNA-binding proteins find their targets? {\it Nucl. Acids Res.} {\bf 32}, 3040-3052 (2004).

\bibitem{mirny09} L. Mirny, M. Slutsky, Z. Wunderlich, A. Tafvizi, J. Leith and A. Kosmrlj,  How a protein searches for its site on DNA: the mechanism of facilitated diffusion. {\it J. Phys. A: Math. Theor.} {\bf 42}, 434019 (2009).

\bibitem{kolomeisky11} A.B. Kolomeisky, Physics of protein-DNA interactions: Mechanisms of facilitated target search. {\it Phys. Chem. Chem. Phys.} {\bf 13} 2088-2095 (2011).

\bibitem{veksler13} A. Veksler and A.B. Kolomeisky, Speed-selectivity paradox in the protein search for targets on DNA: Is it real or not? {\it J. Phys. Chem. B} {\bf 117}, 12695-12701 (2013).

\bibitem{lange15a} M. Lange, M. Kochugaeva and A.B. Kolomeisky, Protein search for multiple targets on DNA. {\it J. Chem. Phys.} {\bf 143}, 105102 (2015).

\bibitem{lange15b}  M. Lange, M. Kochugaeva and A.B. Kolomeisky, Dynamics of the protein search for targets on DNA in the presence of traps. {\it J. Phys. Chem. B} {\bf 119}, 12410-12416 (2015).

\bibitem{shvets15} A.A. Shvets and A.B. Kolomeisky, Sequence heterogeneity accelerates protein search for targets on DNA. {\it J. Chem. Phys.} {\bf 143}, 245101 (2015).

\bibitem{shvets16} A.A. Shvets and A.B. Kolomeisky, Crowding on DNA in protein search for targets. {\it J. Phys. Chem. Lett.} {\bf 7}, 2502-2506 (2016).

\bibitem{shvets15b} A.A. Shvets, M. Kochugaeva and A.B. Kolomeisky, The role of static and dynamic obstacles in the protein search for targets on DNA. {\it J. Phys. Chem. B} {\bf 120}, 5802-5809 (2015).

\bibitem{shvets16b} A.A. Shvets and A.B. Kolomeisky, The role of DNA looping in the search for specific targtes on DNA by multisite proteins. {\it J. Phys. Chem. Lett.} {\bf 7}, 5022-5027 (2016).

\bibitem{kochugaeva16} M.P. Kochugaeva, A.A. Shvets and A.B. Kolomeisky, How conformational dynamics influences the protein search for targets on DNA. {\it J. Phys. A: Math. Theor.} {\bf 49}, 444004 (2016).

\bibitem{shin18} J. Shin and A.B. Kolomeisky, Surface-assisted dynamic search processes. {\it J. Phys. Chem. B} {\bf 122}, 2243-2250.

\bibitem{esadze14} A. Esadze, C.A. Kemme, A.B. Kolomeisky and J. Iwahara, Positive and negative impacts of nonspecific sites during targte location by a sequence-specific DNA-binding protein: Origin of the optimal search at physiological ionic strength. {\it Nucl. Acids Res.} {\bf 42}, 7039-7046 (2014).

\bibitem{kochugaeva17} M.P. Kochugaeva, A.A. Shvets and A.B. Kolomeisky, On the mechanism of homology search by ReacA protein filaments. {\it Biophys. J.} {\bf 112}, 859-867 (2017).

\bibitem{shvets17} A.A. Shvets and A.B. Kolomeisky, Mechanism of genome interrogation: How CRISPR RNA-guided Cas9 proteins locate specific targets on DNA. {\it Biophys. J.} {\bf 112}, 1416-1424 (2017).



\end{thebibliography}

\end{document}